# Subwave spikes of the orbital angular momentum of the vortex-beams in a uniaxial crystal

T. Fadeyeva, C. Alexeyev, A. Rubass, A. Zinov'ev , V. Konovalenko and A. Volyar*

*Department of Physics, Taurida National V.I. Vernadsky University Vernadsky av.4, Simferopol, Ukraine, 95007*
*Corresponding author: volyar@crimea.edu*

*We have theoretically predicted the gigantic spikes of the orbital angular momentum caused by the conversion processes of the centered optical vortex in the circularly polarized components of the elliptic vortex beam propagating perpendicular to the crystal optical axis. We have experimentally observed the conversion process inside the subwave deviations of the crystal length. We have found that the total orbital angular momentum of the wave beam is conserved.*
OCIS Codes: 350.5030, 260.6042, 260.1180, 260.0260

It is well known that propagation of the circularly polarized plane wave perpendicular to the optical axis of a birefringent uniaxial crystal is accompanied by the complete energy transport between its circularly polarized components [1]. More complex process is observed in the paraxial beams. The initial paraxial beam with a circular cross-section splits into ordinary and extraordinary beams inside the crystal. The ordinary beam propagates without any structural transformations while the extraordinary one becomes elliptically deformed [2,3]. Such an elliptical deformation is explained by different scales along the transverse and directions in the extraordinary beam. Although the value of the elliptical deformation is rather small it has an appreciable impact upon the propagation of vortex-beams resulting e.g. in precession of the off-axis optical vortices [4] in the rotating crystals. The polarization and the wavefront transformations of the beam cannot but affect the angular momentum of the beam as a sum of the orbital (OAM) and spin (SAM) angular momenta and also the crystal rotation. In the general case, we must say about the conservation of a superposition of the OAM, SAM and the mechanical angular momentum along the propagation axis exerted by the wave beam in uniaxial and in biaxial crystals [5,6,7]. In the recent paper [8] it was shown that differences in wavefront structures of the ordinary and extraordinary generalized Laguerre-Gaussian beams in a uniaxial crystal make the centered optical vortices evolve in complex topological ways.

The aim of our paper is to analyze theoretically and experimentally the conversion of the centered optical vortices in the singular beam propagating perpendicular to the optical axis of a uniaxial crystal in terms of the evolution of the OAM and SAM.

1. We consider the propagation of the monochromatic circularly polarized elliptic singular beam bearing singly charged optical vortex perpendicular to the optical axis of a uniaxial birefringent medium with the permittivity tensor written in the diagonal form:   (see Fig.1). The electric field of the beam can be presented as a superposition of the ordinary and extraordinary and beams with the linear polarizations directed along the and axes, respectively [8].

$$E_+ = E_x + E_y, \quad E_- = E_x - E_y, \quad (1)$$

where

$$E_x = \left(\frac{x}{w_x \sigma_{xx}} + i\frac{y}{w_y \sigma_{xy}}\right) \times$$
$$\times \exp\left\{-\frac{x^2}{w_x^2 \sigma_{xx}} - \frac{y^2}{w_y^2 \sigma_{xy}} - ik_x z\right\} / \sqrt{\sigma_{xx}\sigma_{xy}}, \quad (2)$$

$$E_y = \left(\frac{x}{w_x \sigma_{yx}} + i\frac{y}{w_y \sigma_{yy}}\right) \times$$
$$\times \exp\left\{-\frac{x^2}{w_x^2 \sigma_{yx}} - \frac{y^2}{w_y^2 \sigma_{yy}} - ik_y z\right\} / \sqrt{\sigma_{yx}\sigma_{yy}}, \quad (3)$$

$\sigma_{xx} = 1 + iz/z_{xx}$, $\sigma_{xy} = 1 + iz/z_{xy}$, $k_x = k_0 n_x$, $k_y = k_0 n_y$, $z_{xx} = k_x w_x^2/2$, $z_{xy} = k_x w_y^2/2$, $\sigma_{yx} = 1 + iz/z_{yx}$, $\sigma_{yy} = 1 + iz/z_{yy}$, $z_{yx} = k_y w_x^2/2$, $z_{yy} = (k_x^2/k_y)w_y^2/2$, $n_x = \sqrt{\varepsilon_1}$, $n_y = \sqrt{\varepsilon_2}$, $k_0$ is the wavenumber in free space.

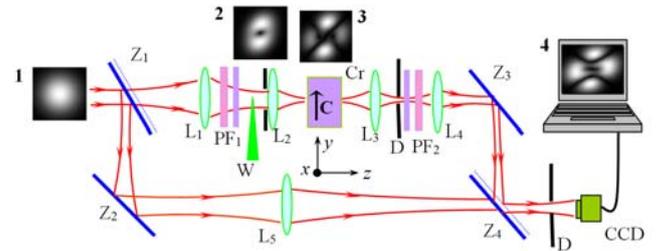

Fig.1 (Color online) Sketch of the experimental set-up: Z1,2,3,3 – interferometer, L – lenses, PF – polarization filters, D – diaphragms, W – optical wedge, Cr – SiO2 crystal, C – optical axis direction ∶ 1,2,3,4 – intensity distributions of RHP component.

The wave beam (1) has only the right hand circularly polarized component (RHP) $E_+$ at the $z=0$ plane. The major axes of its elliptic cross section is directed along the $x-$ and $y-$ axes of the referent frame so that $w_x$ and $w_y$ stand for the beam waists. The optical vortices in the initial beam components $E_x$ and $E_y$ have the same ellipticity of their cores. As the elliptic vortex beam propagates through the crystal its wave structure changes right along. To trace this process it is convenient to make some approximations in eqs (1-3) considering the beam field near the axis. We assume that the birefringence in the crystal is rather small; besides, the ellipticity of the initial beam is small too. In the paraxial approximation for small distances $z/z_{xx}, z/z_{yy} \ll 1$ we can account that $\sqrt{\sigma_{xx}\sigma_{xy}} \approx \sqrt{\sigma_{yx}\sigma_{yy}}$ and $|\sigma_{xx}| \approx |\sigma_{xy}|, |\sigma_{yx}| \approx |\sigma_{yy}|$ in the $E_+$ component. Using eqs (2) and (3) we obtain:

$$E_+ \sim \frac{x}{w_x}\cos(\delta\phi_x) + i\frac{y}{w_y}\cos(\delta\phi_y)\exp(-i\delta\beta\, z), \qquad (4)$$

where $2\delta\phi_x = \arg(\sigma_{xx}) - \arg(\sigma_{yx}) + \Delta\beta\, z$,
$2\delta\phi_y = \arg(\sigma_{xy}) - \arg(\sigma_{yy}) + \Delta\beta\, z$, $2\Delta\beta = k_x - k_y$,
$2\delta\beta = \arg(\sigma_{xx}) - \arg(\sigma_{yx}) + \arg(\sigma_{yy}) - \arg(\sigma_{xy})$.

The beam (4) has characteristic features of the astigmatic beam. Similar to the wave beams in the astigmatic lens system [9] the centered optical vortices in eq. (4) can change the sign of the topological charge. The value $\Delta\beta$ is very large $\Delta\beta \sim 10^6 m^{-1}$ whereas the value $\delta\beta \sim 0.1 m^{-1}$ cannot affect the relation between the terms in eq. (4) so that the major contribution to the conversion process is made by the oscillations of $\cos(\delta\phi_x)$ and $\cos(\delta\phi_y)$. The conversion of the optical vortex sign in eq.(4) occurs where $\cos(\delta\phi_x)$ and $\cos(\delta\phi_y)$ have different signs, e.g. $0 < \delta\phi_x(z_1) < \pi/2$, $3\pi/2 < \delta\phi_x(z_1) < 2\pi$, $\pi/2 < \delta\phi_y(z_2) < 3\pi/2$ from whence we obtain the estimates for the deviations of the crystal length $\Delta z = z_2 - z_1$ in the range $\Delta z \sim e^2/\left[(n_y - n_x)k_0^2 z_{xy}\right]$, where $e$ stands for the excentricity of the initial beam cross-section. Thus, for the $SiO_2$ crystal: $n_x = 1,54282$, $n_x = 1,55188$ at the wavelength $\lambda = 0.63\,\mu m$ and the beam ellipticity $w_y/w_x = 0.5$ we find the range of the crystal lengths, inside which the conversion process can be observed: $\Delta z \sim 0.1\lambda$ that extends inside the subwave variations of the crystal length. The range $\Delta z$ increases when growing the excentricity $e$ and comes up to zero for the circular initial cross-section. The vortex conversion occurs in vicinity of the crystal length $\bar{z} = (2m+1)\pi/(k_x - k_y)$, $m = 0,1,2,\ldots$, when the major portion of the wave energy is transported from the RHP $E_+$ into the left hand polarized (LHP) $E_-$ component. Typical computer simulated intensity distributions of the RHP components corresponding to the first stage of the conversion process are illustrated in Fig 1 (**2,3,4**) for the $SiO_2$ crystal, $w_x = 10\,\mu m$, $w_y = 15\,\mu m$.

The conversion of the centered optical vortex in one of the circularly polarized beam component is undoubtedly connected with the radical transformation of the OAM. A simple algebra in eqs (1)-(3) shows that the total OAM $l_z$ of the beam per photon as a sum of the OAM $l_z^{(+)}$ and $l_z^{(-)}$ in the beam components is conserved over the crystal length:

$$l_z = l_z^{(x)} + l_z^{(y)} = \left(w_x/w_y + w_y/w_x\right)/2 \geq 1, \qquad (5)$$

$l_z^{(x)}$ and $l_z^{(y)}$ are the OAM of the partial beams (2) and (3). Consequently, the sum of the SAM $S_z$ of the beam and mechanical angular momentum $M_z$ exerted by the beam into the crystal is conserved too: $S_z + M_z = const$. The above equation shows that the elliptic vortex-beam gets an excessive OAM associated with the additional inclination of the optical currents relative to the beam axis [10]. Fig. 2 illustrates transformations of the OAM in the RHP and LHP components accompanied by gigantic spikes of the OAM amplitude. The average width of the spikes is about $\Delta z = 0.15\,\mu m$. The gigantic spike in one of the polarized component is compensated by a slight alteration of the OAM in the other polarized component so that the total OAM stays put. The SAM oscillates synchronically with transformations of the handedness of the circular polarization in the beam from +1 to -1 and vice versa at the beating length $\delta z = 2\pi/(k_x - k_y)$.

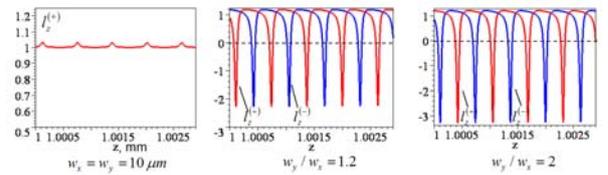

Fig.2 (Color online) Subwave spikes of the OAM $l_z^{(\pm)}$ in the RHP and LHP components of the vortex beam in the $SiO_2$ crystal for different ellipticities $w_y/w_x$

**2.** In order to observe the conversion of the centered vortex we used the experimental set-up shown in Fig.1 (the detailed description of the set-up major units can be found, e.g., in [4, 11]). The $SiO_2$ crystal with the length $1cm$ was placed into the thermostat and heated up to $30\,^oC$. The temperature is measured when cooling the crystal down the room temperature accurate within $0.01\,^oC$. The experimental set-up enabled us to measure not only the intensity distributions and interference patterns but also the polarization distributions. The elliptic vortex beam of a positive topological charge $l=+1$ is shaped by the optical wedge W. The ellipticity of the beam cross-section is about $w_y/w_x \approx 0.61$ whereas the beam waist $w_x \approx 10\,\mu m$ at the crystal input. We measured the positions of the interferential forks associated with optical vortices in the RHP beam component and positions of the polarization singularities. After computer processing the data we plotted the vortex trajectories in the reference frame – the transverse coordinate $x, y\,\mu m$ and the temperature $t\,^oC$. We cannot use the longitudinal coordinate $z$ because there are three processes responsible for the beam transformations via the temperature, namely, the thermal dilatation with the coefficient $\alpha_z = 13.2 \times 10^{-6}\,^oC^{-1}$, the refractive index transformation with the coefficients $\alpha_o = -5\times10^{-6}\,^oC^{-1}$ and $\alpha_e = -6\times10^{-6}\,^oC^{-1}$ and also the photoelasticity effect. All the processes have the same order of magnitude of the corresponding coefficients and can partially compensate each other. At the same time, we can compute the effective deviation of the crystal optical length $s=(n_y-n_x)z$. The vortex trajectories for the RHP component are shown in Fig.3. The conversion process in the figure is accompanied by the corresponding intensity distributions near the critical planes. In vicinity of the plane $z=\pi(2m+1)/(k_x-k_y)$, two pairs of vortex dipoles are born (points 1 and 2 in Fig.3).

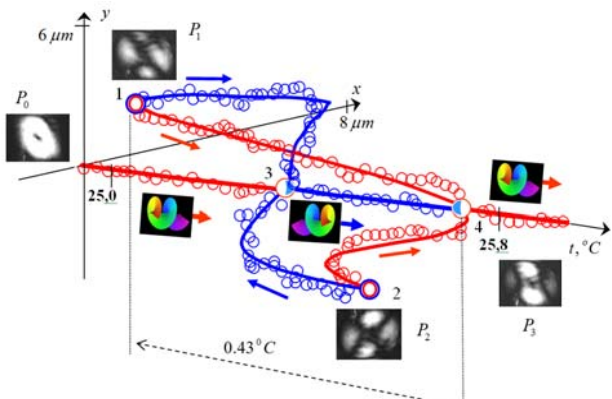

Fig.3 (Color online) Vortex trajectories of $E_+$ component in vicinity of the conversion process. circles – vortex positions, solid lines – theory, points 1, 2 – birth of the vortex dipole, 3, 4 – annihilations of vortices, P_0 – the beam intensity at the crystal input, P1, P2, P3 – intensities at the 1,2, 4 points, respectively

Two negatively charged vortices from each pair come near to the axis and annihilate with the centered vortex at the point 3 so that the negatively charged vortex moves along the beam axis. Then two off-axis positively charged vortices come near the axis and annihilate with the centered negatively charged vortex at the point 4 so that the positively charged vortex resumes its motion along the beam axis. The temperature range of the vortex conversion is about $\Delta t \approx 0.43\,^oC$. This corresponds to the change of the optical length $\Delta s = 0.0088\,\mu m$. It is inside such a subwave deviation of the crystal length that the gigantic spikes of the OAM occur.